\documentstyle[12pt]{article}

\newcommand{\e}[1]{\label{eq:#1}}
\newcommand{\ee}[1]{(\ref{eq:#1})}

\newcommand{\eq}{\begin{equation}}
\newcommand{\eqe}{\end{equation}}
\newcommand{\eqa}{\begin{eqnarray}}
\newcommand{\eqae}{\end{eqnarray}}

\newcommand{\del}{\partial}

\newcommand{\ba}{\mbox{\boldmath $a$}}
\newcommand{\bk}{\mbox{\boldmath $k$}}
\newcommand{\bq}{\mbox{\boldmath $q$}}
\newcommand{\bv}{\mbox{\boldmath $v$}}
\newcommand{\bl}{\mbox{\boldmath $l$}}
\newcommand{\bE}{\mbox{\boldmath $E$}}

\begin{document}

\pagestyle{empty}
\hfill{NSF-ITP-93-33}

\hfill{UTTG-09-93}
\vspace{24pt}

\begin{center}
{\bf Low Energy Dynamics of the Spinon-Gauge System}

\vspace{24pt}
\renewcommand{\thefootnote}{\fnsymbol{footnote}}
Joseph Polchinski\footnote{joep@sbitp.ucsb.edu}
\renewcommand{\thefootnote}{\arabic{footnote}}
\addtocounter{footnote}{-1}

\vspace{24pt}
\sl

Institute for Theoretical Physics \\ University of California \\
Santa Barbara, California 93106-4030

\vspace{12pt}
\rm and\\
\vspace{12pt}
\sl
Theory Group\\ Department of Physics \\ University of Texas
\\ Austin, Texas 78712
\rm

\vspace{24pt}
{\bf ABSTRACT}

\end{center}
\baselineskip=18pt

\begin{minipage}{4.8in}
The effective field theory describing the normal phase of the high-$T_c$
cuprates is evidently not
the usual Fermi liquid theory.  It has been proposed that it must include
a dynamically generated gauge field.  Even
the simplest such theory, with spinon and gauge fields only,
has complicated dynamics, becoming strongly coupled at low energy.
We show that in a large-$n$ approximation the theory can be solved and has a
nontrivial fixed point.  Also, we find that there is no antiferromagnetic
instability at weak coupling.
\end{minipage}

\vfill

\pagebreak
\baselineskip=21pt
\pagestyle{plain}
\setcounter{page}{1}

\renewcommand{\thefootnote}{\fnsymbol{footnote}}
\baselineskip=21pt

\section{Introduction}

The high-$T_c$ superconducting cuprates present a great puzzle in quantum
field theory.  For most conductors, the effective field theory at energies
below the electronic scale is the Landau Fermi liquid theory.\footnote{For
discussions of Fermi liquid theory in the language of effective field theory
see refs.~[1] and [2]; related work appears in refs.~[3] and [4].}
The normal phase
of the cuprates, however, is not described by this effective theory; various
quantities have the wrong energy dependence.  For example, the decay rate
$\Gamma$ of a current carrier in a Fermi liquid (in two or more
dimensions) goes as the square of the excitation energy $E$ or the temperature
$T$, whichever is larger.  In the normal phase of the cuprates, $\Gamma$ is
linear in $E$ or $T$.  Thus, it appears that interactions at low energy are
enhanced relative to those in a Fermi liquid.\footnote{See, for example,
the contributions of Anderson and Lee in ref.~[5], and the discussion between
Anderson and Schrieffer in ref.~[6].  A recent overview of theoretical ideas
and experimental results can be found in ref.~[7].}

It is possible to obtain enhanced low energy interactions in a Fermi liquid if
the Fermi surface is special, with nesting and/or van Hove singularities.
Such a theory is fine-tuned, however: the shape of the Fermi surface is
relevant in the renormalization group sense.  Any perturbation which changes
the shape of the Fermi surface produces an effective infrared cutoff on the
enhanced low energy behavior.  The generality of the anomalous normal state
behavior, and its stability against changes in doping, argue strongly against
such a fine-tuned explanation.  Most striking is Bi2201, a cuprate with the
low transition temperature of 7K.  The anomalous behavior is observed right
down to the transition, or less than $10^{-3}$ times the electronic scale
$E_0$ (which is roughly 1 eV or 10$^4$ K) and
is stable against changes in doping of order 10\%[8].

We therefore work with the assumption that the low energy theory must be {\it
natural}.  That is, we seek {\it an effective field theory of a conductor with
non-Fermi liquid behavior that is stable against changes in any relevant
parameters}.  Although simply stated, this is in fact a very tall order, and
it is not clear that {\it any} of the vast number of proposed theories
satisfies it.  Needless to say, this subject is not without controversy, and
one can find opinions on all sides of the issue.
The reader might wonder whether some small parameter might enter so as to
lessen
the actual fine tuning.  The author is of course not an expert in condensed
matter physics, and can only remark that changes in doping (in materials
without CuO chains to absorb carriers) affect the electronic properties in
the CuO$_2$ planes quite directly---that is, they should affect the parameters
in the effective action without suppression.

The list of fields which may appear naturally in a low energy effective
theory is very short.  Fermions with a Fermi surface are natural (again, see
refs.~[1,2] for a discussion from the effective field point of view).  The
Fermi
liquid is the general effective field theory of such fermions.  There is
ample evidence that the cuprates have a Fermi surface, so the non-Fermi
liquid behavior is evidently due to interaction with additional fields.
Scalars are natural if they are Goldstone bosons.  For example, the low
energy theory will always include the phonon field.  Below the Debye
temperature this contributes to $\Gamma$ only as max$(E,T)^3$ and so is not
the source of the non-Fermi liquid behavior, particularly in  Bi2201.  No
other continuous symmetries are broken in the cuprates at the dopings of
interest.  It is notable that the cuprates have in general an
antiferromagnetic phase, whose low energy fluctations will include spin
waves.  Near this phase, if the transition is second order, the gap for
scalar spin fluctuations will still be small.  However, the
superconductivity and non-Fermi liquid behavior occur at dopings 10\% to 30\%
away from the antiferromagnetic phase, so the gap should be of order
$0.1^{0.7}$ to $0.3^{0.7} E_0$,\footnote{The exponent $0.7$ is just the
reciprocal of the dimension 1.4 of the relevant perturbation, the scalar
mass-squared.} and these fluctuations are irrelevant at the $10^{-3} E_0$ of
Bi2201.

This would appear to leave one other possibility, a gauge field.  The
electromagnetic field, however, is not the source of the non-Fermi liquid
behavior.  The scalar potential is shielded and short ranged, and so only
contributes to the usual four-fermi interaction of the Fermi liquid.  The
vector potential is not shielded but its effects are suppressed by $c^2$.  It
does in fact lead to non-Fermi liquid behavior[9], but only at energies much
lower than those of interest.

It appears that a second gauge field, generated from the dynamics of the
underlying electrons, is necessary.  This is not impossible, and  may
actually be rather plausible.  In the Fermi liquid theory, the low energy
fields are essentially the same as the short distance fields.  This is not
surprising if the interactions are in some sense weak, but it need not always
be the case.  Starting from models of strongly interacting electrons, it has
been argued that the resulting low-energy theory may indeed include a
dynamical gauge field.\footnote{For a review of recent work in this area,
see ref.~[10].}  There is still a major bifurcation, depending on
symmetry.  If $P$ and $T$ are spontaneously broken, a Chern-Simons term for
the dynamical gauge field is allowed and is the most relevant term in the
gauge action.  If $P$ and/or $T$ is unbroken, the Chern-Simons term is
forbidden and the Maxwell action is the most relevant.
\addtocounter{footnote}{-4}

Each possibility has received attention, but it does seem that the $P$ and
$T$ violating case has received much more study in the field theory
literature.  Given that experiments disfavor $P$ and $T$ violation, even with
a sign that alternates between planes[11], and that the $P$ and $T$ invariant
theory is interesting and nontrivial, it is worthwhile to explore the latter
further.  Non-Fermi liquid behavior in the $P$ and $T$ conserving gauge theory
has been discussed in refs.~[12-15].

We will not in this paper attempt to derive a low energy gauge theory from the
underlying dynamics.  Rather, in the spirit of effective field theory, we
will start with a plausible set of low energy fields and symmetries, write
the most general effective Lagrangian, and analyze the resulting physics.
For completeness, though, let us now give a brief flavor of the arguments
which lead to a dynamical gauge field[16,17].  Start with electrons moving
on a
lattice, with a strong repulsion forbidding two electrons on a site.  It is
useful to replace this inequality constraint with an equality, regarding each
site as occupied either by a spin-up or spin-down electron, or by a hole.  In
terms of the {\it spinon} field $f_{i\alpha}$ and {\it holon} field $b_i$,
\eq
\sum_{\alpha=1}^2 f_{i\alpha}^\dagger f_{i\alpha} + b_i^\dagger b_i = 1
\eqe
for each site $i$.  If the average density of electrons per site is
$1-x$, the spinon density is $1-x$ and the holon density $x$.  The electron
field $\psi_{i\alpha}$ destroys a spin and creates a hole, so
\eq
\psi_{i\alpha} = f_{i\alpha} b_i^\dagger.    \e{electron}
\eqe
Notice that
the spinon-holon description is redundant, in the sense that  a local phase
redefinition $f_{i\alpha}(t) \to e^{i \lambda_i(t)} f_{i\alpha}(t)$, $b_i(t)
\to e^{i \lambda_i(t)} b_i(t)$ leaves the original field~\ee{electron}
unchanged.  This redundancy can be promoted to a dynamical gauge symmetry,
just as occurs in the $CP(n)$ sigma model.  In particular, replace $SU(2)$
spin with $SU(n)$ and integrate out the spinons to find
the large-$n$ action for the singlet mean
field $\Delta_{ij} = f_{i\alpha}^\dagger f_{j \alpha}$[17].
One finds that
for some ranges of the various hopping and spin interactions the system will
be
in the {\it uniform phase}, where
the large-$n$ mean field has $\Delta_{ij}$ non-zero and equal (up to
gauge equivalence) on every link.  In this phase, the fluctuations are
described by independently-propagating spinon and holon fields, and by the
gauge field $a_{ij}$ which is just the phase of $\Delta_{ij}$. Again, we will
not try to critically evaluate this reasoning, except to remark that we see
no objection in principle.

In this paper, we will consider just the spinon-gauge system, which is already
quite nontrivial.  This would correspond to $x=0$, half-filling, where a state
of a hole and anti-hole pays a large Coulomb energy
and does not appear in the low energy theory.
In the spinon-gauge system, the
gauge interaction is {\it relevant}, growing at low energy[18].  Understanding
this strongly coupled system is our principle goal.
In the next section we show, using the same large-$n$ approximation as above
to
organize the perturbation theory, that the growth of the coupling is
effectively cut off by quantum effects, leading to a nontrivial fixed point.
The key idea is Migdal's  theorem, which we show to be valid here in the
large-$n$ limit.  This makes it possible to derive closed integral equations
for the gauge and spinon propagators, which are easily solved due to the
kinematics. The result is consistent with
various conjectures and unpublished remarks in the literature, but we know of
no
derivation within a systematic approximation.
We go on to consider possible instabilities of the resulting state.
We find that there is no instability to formation of a spin or charge
density wave at weak coupling (large $n$); at small $n$ we cannot calculate
reliably but find no positive sign of symmetry breaking.  We also point out
an instability to development of a $P$ and $T$ violating gauge field
when the spinon Fermi surface is near a van Hove singularity.
We work at zero temperature throughout.  The extension to finite temperature
brings in interesting new issues[19].

We conclude this introduction with a few remarks about the holons.  Under the
electromagnetic and dynamical gauge symmetries, the charge assignments are
$f_{i\alpha}: (0, 1)$ and $b_i : (1, 1)$.  Only the difference of the
electromagnetic charges of the spinon and holon is physical, as it must equal
the charge of the electron.  The separate charges may be changed by shifting
the dynamical field $a_\mu$ by a constant times the electromagnetic field
$A_\mu$.  Note, too, that the holons are a necessary part of any theory of the
normal state: if they are absent, no fields in the low energy theory carry
electromagnetic charge, and we do not have a conductor.  The holons, however,
present a severe naturalness problem.  There are two obvious relevant terms in
any low energy effective Lagrangian, namely
\eq
\mu_f \sum_{i,\alpha}
f_{i\alpha}^\dagger f_{i\alpha} \qquad {\rm and} \qquad \mu_b \sum_i
b_i^\dagger b_i.
\eqe
Here $\mu_f$ and $\mu_b$ are chemical potentials, whose
values are fixed by the spinon density $1-x$ and the holon density $x$.  The
spinon chemical potential does not render the low energy theory unnatural: the
existence of a Fermi surface is stable against changes in $\mu_f$.  The holon
chemical potential, resembling as it does the a scalar mass term (the fatal
flaw of the Standard Model), is dangerous.  At very low holon densities
$\mu_b$ is small, but at the dopings of order 10\% which are of interest, it
is only slightly below the electronic scale.  A non-zero density of bosons
will tend to condense, and the characteristic temperature is again only
slightly below the electronic scale.\footnote{If the holons are
two-dimensional there is no sharp transition at nonzero temperature, but at
low temperature the electromagnetic response will resemble that of a
bose-condensed system to exponential accuracy.}  The holon and dynamical
gauge boson become massive due to this spontaneous breaking, leaving only the
spinons in the low energy theory (now effectively carrying electric charge,
because the massless field is $A_\mu - a_\mu$), and we are back in the Fermi
liquid theory.

If this theory is in fact to explain the normal state of the cuprates, it is
necessary to find a phase in which the holons conduct without
superconducting.  Perhaps the fluctuations of the dynamical gauge field
prevent the tendency toward order.  Arguments in this direction are made in
refs.~[12,20], but it is difficult to see an effect sufficiently strong to
provide the orders of magnitude seen in Bi2201.
So it is not at all clear that this theory solves the naturalness problem
posed; a much better understanding of the dynamics of the holons is needed.
In
any case we will study the spinon-gauge sector as an interesting exercise in
field theory.  Incidentally, the $P$ and $T$ violating theory seems to fare no
better: the natural scale for superconductivity is again the electronic
scale, and it is not clear how a normal state can survive down to much lower
temperatures.  We should also mention Anderson's idea, which leads to an
effectively nonlocal four-fermi interaction without introducing an associated
gauge field in the low energy theory[21].  Certainly it is possible in
principle
to obtain enhanced infrared interactions and non-Fermi liquid behavior in
such an effective theory; the question is whether such a nonlocal interaction
can actually arise from the underlying dynamics.  It is not clear, within the
effective field interpretation of Fermi liquid theory, how it can do so[1].

\section{Strong Coupling and a Nontrivial Fixed Point}

The gauge invariant kinetic terms for the spinons and dynamical gauge field
take the form
\eqa
L &=& \int \frac{d^2 \bk}{(2\pi)^2}
\, f^\dagger_\alpha( \bk ) \Bigl\{ i\del_t - a_0 - {\cal E}( \bk
 ) + \mu_f \Bigr\} f_\alpha( {\bk})
\nonumber\\[2pt]
&&\quad + \quad \int \frac{d^2 \bk\, d^2 \bq}{(2\pi)^4}
\, f^\dagger_\alpha( \bk + \bq) f_\alpha( {\bk}) \ba(\bq) \cdot \frac{\del}
{\del \bk} {\cal E}( \bk + \bq/2) + \ldots
\nonumber\\[2pt]
&&\quad + \quad \frac{1}{2} \int \frac{d^2 \bq}{(2\pi)^2}\,
\Bigl\{ \epsilon_0 \bE(\bq) \cdot \bE(-\bq) - \eta_0 B(\bq) B(-\bq)  \Bigr\}.
\eqae
The gauge field is presumed to be disordered in the direction perpendicular to
the CuO$_2$ planes, so the problem is two-dimensional.  Here, ${\cal E}(\bk)$
is
the single-particle spinon energy, $\bE$ and $B$ are the field strengths
for the
dynamical potential $(a_0, {\ba})$, and $\epsilon_0$ and $\eta_0$ are
parameters.  Other terms will be irrelevant.  In particular, since gauge field
momenta of interest will be $q << k_F$, higher terms in the gauge Lagrangian
are irrelevant.

Exchange of a gauge boson $(\omega, {\bq})$ yields a four-fermi
interaction
\eq
V \propto \frac{1}{\epsilon_0 \omega^2 - \eta_0 q^2}. \e{4f}
\eqe
For given $\omega$, the dominant momenta are then $O(\omega/v)$, where
$v = ( \eta_0 / \epsilon_0 )^{1/2}$ is of order the Fermi velocity.
The overall interaction (including a factor of $q$ from the volume of momentum
integration) goes as $\omega^{-1}$.  Since the four fermi interaction with
{\it constant} coefficient is marginal for $v q \sim \omega$[1,2], the
interaction~\ee{4f} is {\it relevant}, growing strong as $\omega \to 0$.

It is this strongly coupled theory
that we wish to understand.  When a coupling is relevant in field theory,
there are two broad possibilities for the low energy dynamics.  The first is
that something interesting happens---bound states, symmetry breaking---and the
low energy spectrum bears little resemblance to the quanta of the original
Lagrangian.  It is then necessary to start over again, identifying the new
effective theory which describes the actual low energy spectrum.  The second
possibility is a nontrivial fixed point where the quantum effects cut off the
growth of the coupling.  In this case, the low energy fluctuations, although
not free, still correspond to the fields in the Lagrangian.  In the present
case we shall argue, making use of a large-$n$ expansion to control the
perturbation theory, that the latter occurs.

To start, the estimate~\ee{4f} is inaccurate for a reason that is well-known.
Over much of $(\omega,\bq)$-space the fluctuations
of the gauge field are controlled not by the classical action but by the
effective action from the fluctuations of the fermions.  In order to
make a systematic treatment, we at this point take $SU(n)$ spinons and make
the
large-$n$ approximation.  The leading-$n$ effective action comes only
from one loop (the random phase approximation),\footnote
{We have scaled the tree-level action with $n$ as well;
this is convenient but inessential.}
\eq
L_{\rm gauge} \sim \frac{n}{4} \int \frac{d^2{\bq}}{(2\pi)^2}\, \Biggl(
\Bigl\{ \epsilon_0 + \epsilon_1 (\omega,{\bq}) \Bigr\}
\bE(\bq) \cdot \bE(-\bq) - \Bigl\{ \eta_0 + \eta_1 (\omega,{\bq})
\Bigr\} B(\bq) B(-\bq) \Biggr).
\eqe
For $\omega < qv << k_F v$, the RPA correction is known to have the form
\eqa
\epsilon_1 (\omega,{\bf q}) &\sim& \frac{\epsilon_0}{q^2 \ell_D^2}
\nonumber\\[2pt]
\eta_1 (\omega,{\bf q}) &\sim& \chi_f - i \gamma \frac{|\omega|}{q^3} ,
\eqae
where $\ell_D$, $\chi_f$, and $\gamma$ are constants.  Due to the $1/q^2$ in
$\epsilon_1$, the four-fermi interaction from exchange of $a_0$ now
approaches a
finite constant at small $\omega$ and $q$.  This is Debye screening, discussed
in standard texts.  This is then no longer a relevant interaction, but just
a contribution to the usual four-fermi interaction of Landau theory.  The
longitudinal $a_L$ may be set to zero (Coulomb gauge), while the effective
four-fermi interaction from exchange of transverse gauge bosons is
now of order
\eq
\frac{n^{-1}}{(\epsilon_0 + \epsilon_1 )\omega^2 - (\eta_0 + \chi_f) q^2
+ i \gamma |\omega|/q } . \e{gprop}
\eqe
For $q >> k_F^{2/3} (\omega / v)^{1/3}$, the $q^2$ term in the denominator
dominates.\footnote{The reader may find it useful to think in terms of units
$k_F = v = 1$, so that all electronic scales are of order~1.}
For $q << k_F^{2/3} (\omega / v)^{1/3}$, the $\gamma$ term from
Landau damping dominates[22,12].  The interaction is greatest for $q \sim
k_F^{2/3} (\omega / v)^{1/3}$; including a factor of $q$ from the volume of
integration, it is of order $n^{-1} \omega^{-1/3}$.  Although the coupling
grows more slowly than the naive estimate from eq.~\ee{4f}, it is still
relevant and becomes strong at low energy for any fixed $n$[18].  It is this
strong coupling problem that we wish to solve.

The key is Migdal's theorem.  In the above estimates we have treated the
four-fermi operator as {\it marginal}, so that the scaling of the
interaction comes only from the explicit energy-dependence of its
coefficient.  In Fermi liquid theory, however, the four-fermi
operator is {\it irrelevant} at generic kinematics due to Pauli exclusion;
it is suppressed by $\omega / v_F q$.  Applying this factor naively to the
above estimate would give a coupling going as $\omega^{+1/3}$, which is
irrelevant.  This would in turn imply Migdal's theorem, that the gauge vertex
is not radiatively corrected at low energy.\footnote {Again, for a discussion
in the present language, see ref.~[2].}  Of course, we must check this
explicitly in the present case.  The logic of the remainder of the section is
to assume Migdal's theorem, which gives closed Schwinger-Dyson equations
similar to those in strong-coupling superconductivity (Eliashberg theory).
These equations are easily solved due to the special kinematics.  We then go
back and reexamine  the validity of Migdal's theorem for this situation.

Assuming Migdal's theorem, the integral equations for the spinon and gauge
self-energies are
shown graphically in figure~1.  Before giving the explicit form, let us
discuss kinematics.  The momenta of the strongly coupled gauge fields are $q
\sim k_F^{2/3} (\omega / v)^{1/3} >> \omega / v$.  In Fermi liquid theory, the
fermion momenta are of order $\omega / v$ from the Fermi surface, which is
much less than $q$.  The only way a fermion can absorb the much larger $q$
and remain near the Fermi surface is if the gauge momentum is nearly tangent
to the Fermi surface.  In other words, two spinons at generic points on the
Fermi surface, with tangents not parallel, cannot interact strongly.
We thus focus on a single point on the Fermi surface, and
we wish to understand the theory as we scale toward this
point.\footnote{In the next section we consider the strong interaction
between two spinons at distinct points whose tangents are parallel.}
Note that there is still a mismatch: even if the gauge momentum is roughly
parallel to the Fermi surface, due to the curvature of the surface the
distance from the spinon to the surface will generically change by
$O(q^2 / k_F) \sim k_F^{1/3} (\omega / v)^{2/3}$, which is much greater than
the $\omega / v$ of ordinary Fermi liquid theory.  Inspection of diagrams
shows that the former is indeed the region that dominates.

Define ${\bl} = {\bk} -  {\bk}_0$ where ${\bk}_0$ is the point on the Fermi
surface toward which we are scaling.  Rotate the axes so the Fermi surface
runs in the $l_{x}$ direction.
According to the remarks in the previous paragraph, we are interested in the
behavior of amplitudes under the scaling
\eq
\omega \to s \omega, \qquad l_{x} \to s^{2/3} l_{x}, \qquad
l_{y} \to s^{1/3} l_{y} \e{scale}
\eqe
The single-particle energy is
\eq
{\cal E}( {\bk}) = v_F^* l_{y} + \frac{ l_{x}^2}{2m^*}
\e{speng}
\eqe
with higher powers of momentum being irrelevant.  The parameters $v_F^*$
and $m_F^*$ are the effective Fermi velocity and mass at the given point on the
Fermi surface.  The dominant gauge momenta are in the
${x}$-direction, so the strongly coupled transverse gauge field is polarized
in the ${y}$-direction.  The trilinear gauge coupling is then $\del {\cal E}
/ \del l_{y} = v_F^*$, while the quartic interaction $\del^2 E / \del l_{y}^2 =
0$ is irrelevant.

Define the full spinor and transverse gauge propagators
\eqa
<T f^\dagger_{\alpha} (\varepsilon,\bk) f_{\beta} (\varepsilon',\bk') >
&=& (2\pi)^3 \delta(\varepsilon - \varepsilon') \delta^2(\bk - \bk')
\delta_{\alpha\beta} G (\varepsilon, {\bl}) \nonumber\\[2pt]
<T a_{y} (\omega,\bq) a_{y} (\omega',\bq') > &=&
(2\pi)^3 \delta(\omega - \omega') \delta^2(\bq - \bq')
D (\omega, {\bq})        \e{corr}
\eqae
with
\eqa
G(\varepsilon, {\bl}) &=& \frac{i} {\varepsilon e^{i\delta} - v_F^* l_{x} -
l_{y}^2 / 2m^* - \Sigma(\varepsilon, {\bl})} \nonumber\\[2pt]
D (\omega, {\bq}) &=&
\frac{i}{\epsilon_0 \omega^2 e^{i\delta} n/2 - \eta_0 q^2 n/2 - \Pi(\omega,
{\bq}) } . \e{props}
\eqae
The integral equations of figure~2 now take the form
\eqa
-i \Sigma(\varepsilon, {\bl}) = -v_F^{*2} \int \frac{d\omega\, d^2{\bq}}
{(2 \pi)^3}\, D (\omega, {\bq} + \bl) G(\varepsilon - \omega, - \bq)
\e{spineq}
\eqae
and
\eq
-i \Pi(\omega, {\bq}) = nv_F^{*2} \int \frac{d\varepsilon d^2 {\bl}}
{(2 \pi)^3}\, G(\varepsilon, {\bl}) G(\omega + \varepsilon, {\bq} + {\bl}).
\e{geq}
\eqe

To solve these, consider first the spinon self-energy $\Sigma$.  This has
three arguments, $\varepsilon$, $l_{x}$, and $l_{y}$.  However, it can depend
on the momenta only in the form $r = l_{y} + l_{x}^2 / 2m^* v_F^*$, which is
the
distance from the Fermi surface.  One way to see this is to note that,
although we have not assumed rotational symmetry, the Fermi surface is locally
indistinguishable from a round one, where the energy~\ee{speng} would
correspond to a radius $k_F = m^* v_F^*$.  Further, $\Sigma$ does not in fact
depend on $r$ at all.  In eq.~\ee{spineq} the external spinon momentum has
been routed through the gauge line; since the gauge momentum $q$ is much
larger than $r$ under the scaling~\ee{scale}, the result is independent of
$r$.  Thus, $\Sigma$ is a function only of the energy $\varepsilon$.

With this result, the polarization $\Pi$ can now be obtained.  We have just
argued that the momentum dependence of the spinon propagators in eq.~\ee{geq}
is the same as in free field theory.  The momentum integrals can then be
carried out, with result \eqa
\Pi( \omega, {\bq} ) &=& -\frac{in m^*|v_F^*|}{2 q} {\rm sign}
(\omega) \int_{-\omega}^0
\frac{d\varepsilon}{2\pi} \nonumber\\[2pt]
&=& \frac{- i n m^* |v_F^* \omega|}{4\pi q}.
\eqae
\addtocounter{footnote}{-4}
One must sum over all points $j$ where the Fermi surface is tangent to the
${x}$-direction, so the constant $\gamma$ in the Landau damping term of the
gauge propagator~\ee{gprop} is
\eq
\gamma = \frac{1}{2\pi} \sum_j m^*_j |v_{Fj}^* |.
\eqe
The full low-energy
transverse propagator is then
\eq
D (\omega, {\bf q}) = \frac{i (2/n) }{- \chi q^2 + i \gamma
|\omega|/q}, \e{Gres}
\eqe
The coefficient $\chi = \eta_0 + \chi_f$ cannot be determined from
the scaling analysis.  Because it corresponds to renormalization of the local
operator $B^2$, it receives contributions from all parts of the Fermi surface
and all scales.   It must be treated as
an undetermined parameter in the low energy theory (except for one special
circumstance to be discussed in the next section).  The $\omega^2$ term in the
denominator is subleading in the scaling~\ee{scale} and has been dropped.

The integral~\ee{spineq} for the spinon propagator may now be carried out,
again using kinematic simplifications.  Since $q_{y}$ is much less than the
total $q$ of the gauge field, the only strong dependence on $q_{y}$ in the
integrand is in the spinon propagator.  Carrying out the $q_{y}$ integral then
leaves
\eqa
\Sigma(\varepsilon) &=& - \frac{v_F^*}{n} \int \frac{d\omega dq_{x}}
{(2\pi)^2}\, \frac{ {\rm sign}(\varepsilon - \omega)}
{ \chi ( q_{x} + l_x )^2 + \gamma |\omega/(q_{x} + l_x)|}
\nonumber\\[2pt]
&=& -c n^{-1} |\varepsilon|^{2/3} e^{i\pi/6} {\rm sign}(\varepsilon),
\e{sigres}
\eqae
where $c = v_F^* /  \pi 3^{1/2} \gamma^{1/3}
\chi^{2/3}$.
The dependence of the internal spinon propagator on the self-energy has again
dropped out after momentum integration, leaving a simple integral.

In summary, with the
assumption of Migdal's theorem the full gauge and spinon self-energies are
then precisely as would be obtained using the free spinon propagator, even
though the actual spinon propagator is substantially different.\footnote
{This is consistent with conjectures in section VI of ref.~[23]; see also
references therein.}  The results~\ee{Gres} and~\ee{sigres} imply that
under the scaling~\ee{scale} the fields behave simply,
\eqa
f_\alpha(s\varepsilon, s^{1/3} l_x, s^{2/3} l_y) &\sim& s^{-4/3} f_\alpha
(\omega, l_x, l_y) \nonumber\\[2pt]
{\ba}(s\omega, s^{1/3} q_x, s^{2/3} q_y) &\sim& s^{-4/3} {\ba}(\omega,q_x,q_y).
\e{andim}
\eqae
(Note that the term linear in $\varepsilon$ in the denominator of $G$ is
subleading.)  Including a factor of $s^2$ for each $d\varepsilon d{\bk}$, the
gauge interaction~$f^\dagger f {\ba}$ is now {\it marginal}.  This is the main
result.  The interaction of the spinons with the gauge field suppresses the
spinon fluctuations, so that the interaction itself is reduced from relevant
to marginal, and we have a nontrivial fixed point.

It remains to check Migdal's theorem self-consistently.
The dependence of the one-loop vertex correction of figure~2 on scale and $n$
is
\eq
O(\omega^0 n^{-1}). \e{migchek}
\eqe
This can be derived by explicit calculation, but in fact follows at once from
general considerations: the loop graph has the same scaling as the tree level
vertex because the interaction is marginal, and the factor of $n^{-1}$ is
from the gauge propagator.  The significance of the result~\ee{migchek} is as
follows.  Had we found a negative power of energy in the vertex correction, it
would imply that the relevant coupling made Migdal's theorem invalid and our
approach would not work for any fixed $n$.  Had we found a positive power, it
would imply that the exclusion effect was sufficiently strong that Migdal's
theorem works independent of $n$.  As it stands, Migdal's theorem is reliable
for large $n$.  The scaling argument applies to all higher corrections as
well.  Actually, there is the possibility of a logarithm of energy.  This
does not appear in the explicit one loop calculation, and we conjecture that
it does not appear to any order because all fermions are moving in the same
direction.

In summary, we have shown that at large $n$ the relevant gauge interaction
leads to a nontrivial fixed point.  It does seems likely that the same physics
will hold down to $n = 2$, the gauge interaction suppressing the spinon
fluctuations and cutting off the growth of the gauge coupling, but we have no
convincing argument.  A nontrivial fixed point is precisely what is needed to
produce quasiparticle lifetimes $\Gamma \sim E$.  In particular, the pole in
the spinon propagator is at Re($\varepsilon$) = $-$Im($\varepsilon)$ =  $
(v_F^* r / c)^{3/2} 2^{-1/2}$.  Of course, the present results say nothing
about the real materials because we have no holons.

\section{Instabilities}

In the previous section we considered spinons near a single point on the Fermi
surface.  The gauge interaction is also strong between spinons near opposite
points $\bk_0$ and $-\bk_0$, because the tangents are parallel. These spinons
move in opposite directions and so the kinematics allows for a logarithmic
divergence.  In an attractive channel, such a divergence will drive the
marginal interaction relevant and lead to symmetry breakdown.  In particular,
the gauge interaction is attractive between opposite-moving spinon and
hole (which would both have momentum near $\bk_0$), being a magnetic force
between parallel currents[24].  The resulting condensate, having zero charge
and nonzero momentum, would be a charge or spin density wave, depending on the
spin of the condensate.

Consider therefore the one-loop interactions shown in figure~3, where the
incoming and outgoing spinons, and the incoming and outgoing holes, are at
$(\varepsilon, \bk) = (0, \bk_0)$.  The ladder graph is given by
\eqa
&& v_F^{*4} \int \frac{d\varepsilon d^2\bl}{(2\pi)^3}\,
D(\varepsilon, \bl)^2 G(\varepsilon, \bk_0 + \bl)
G(\varepsilon, -\bk_0 + \bl)
\e{ladder}\\[2pt]
&&\quad\qquad =\ -\frac{ 2 i v_F^{*3}}{(2\pi)^2} \int_0^\infty
\frac{d\varepsilon}{\varepsilon^{5/3}}
\int_0^\infty \frac{(2/n)^2 x dx}{( \chi x^3
- i \gamma )^2} \frac{ c n^{-1}e^{i\pi/6} }{ c^2 n^{-2} e^{i\pi/3}
 -  x^4 / 4 m^{*2} }.  \nonumber
\eqae
In the second line we have integrated over $l_y$,\footnote{Notice that for
$\bk = -\bk_0 + \bl$, ${\cal E}( {\bk}) = -v_F^* l_{y} + l_{x}^2 / 2m^*$.}
then
scaled out the energy with $l_x = x |\varepsilon|^{1/3}$.  Similarly, the
crossed ladder is
\eqa
&& v_F^{*4} \int \frac{d\varepsilon d^2\bl}{(2\pi)^3}\,
D(\varepsilon, \bl)^2 G(\varepsilon, \bk_0 + \bl)
G(\varepsilon, -\bk_0 - \bl)
\e{cross}\\[2pt]
&&\qquad\qquad =\ \frac{ 2 i v_F^{*3}}{(2\pi)^2} \int_0^\infty
\frac{d\varepsilon}{\varepsilon^{5/3}} \int_0^\infty
\frac{(2/n)^2 x dx}{( \chi x^3
- i \gamma )^2} \frac{ 1 }{ c n^{-1} e^{i\pi/6} }.  \nonumber
\eqae

We observe a divergence which is not logarithmic but goes rather as
$\varepsilon^{-2/3}$.  In other words, putting the external lines at nonzero
$(\varepsilon_i, \bl_i)$ cuts off the divergence and gives a result
$\varepsilon_1^{-2/3}$ times a function of the scale-invariant ratios
$\varepsilon_i / \varepsilon_1$, $l_{ix} / \varepsilon_1^{1/3}$, and $l_{iy}
/ \varepsilon_1^{2/3}$.  This is the same scaling as the tree-level
interaction~\ee{gprop}, that is, precisely marginal.  Hence, there is
no weak-coupling antiferromagnetic instability.

A logarithmic correction to the scaling of a marginal interaction might have
been expected. The reason that it does not appear is that the marginal
interaction in the antiferromagnetic channel, obtained by exchange of one
gauge field, is nonlocal in time, going as $\omega^{-2/3}$.  High-energy
virtual states can only produce a local effective interaction.  The lowest
dimension local operator is ${\cal O} = f^\dagger_\alpha f_\alpha
f^\dagger_\beta f_\beta$.  This is irrelevant, scaling as $s^{2/3}$---hence
the  $\int d\varepsilon / \varepsilon^{5/3}$ coefficient.

Ref.~[24] identifies a logarithmic divergence in this system.  In that
calculation one gauge propagator has been set to a constant, which does
produce a logarithmically divergent result.  This corresponds to the
anomalous dimension of the operator ${\cal O}$, which is indeed nonzero.
However, a perturbative renormalization of
an irrelevant interaction does not produce an instability.

It is interesting to take this point further.  The one-loop anomalous
dimension for ${\cal O}$ is given by the graphs of figure~4, with 4a and 4b
obtained by omitting one gauge propagator and a factor of $v_F^{*2}$ from the
ladder loop of figure~3a, and 4c and 4d obtained in the same way from the
crossed ladder of figure~3b.  The result is \eq
\alpha = \frac{1}{n} - \frac{4}{3}, \e{anom}
\eqe
which gives the operator an overall scaling $s^{-2 + 1/n}$.  The result is
curiously simple, and does not depend on any parameters other than $n$, in
particular not on the parameters of the Fermi surface.  The signs in
eq.~\ee{anom} are readily understood.  The first term is from the
antiferromagnetic channel (between a spinon and hole of nonzero total
momentum), where the gauge interaction is attractive, while the second is from
the BCS channel (under crossing the gauge line connects a spinon pair of zero
total momentum), where it is repulsive.  The magnitude of the latter term is
notable: it is not suppressed by $n$. The reason is essentially kinematic:
inspection of the integrals~\ee{ladder} and~\ee{cross} shows that the
difference is due to the absence of the $x^4$ term in the last denominator of
the latter. This enhancement corresponds to the kinematic enhancement of the
BCS interaction in ordinary Fermi liquid theory. The presence of a term of
order~1 in the one-loop result~\ee{anom} might appear to signal a breakdown
of perturbation theory.  However, the explanation given above for the
enhancement would imply that it occurs only in BCS ladder graphs, and so
(like the beta function of BCS theory) the one loop result for the $O(n^0)$
anomalous dimension is exact.

One mechanism for production of a spin or charge density wave is for the
interaction~${\cal O}$ to become relevant. With the scaling $s^{-2 + 1/n}$,
${\cal O}$ is irrelevant for all $n \geq 1$, and so we find no indication of
an instability toward antiferromagnetism from this mechanism.  Of course, the
calculation may not be reliable at small $n$.\footnote{Even the $1/n$
correction may be incomplete, as an additional contribution from two loops in
the BCS channel is possible.}

There is a possible instability of another sort. If the coefficient $\chi$ in
the gauge field propagator~\ee{Gres} is negative then the phase we are
considering, the uniform phase, is unstable toward development of a nonzero
value of the dynamical magnetic field, breaking the symmetries $P$ and $T$.
Since $\chi$ gets a contribution from the $B^2$ term in the low energy
effective action, its value is determined in part by virtual high-energy
effects and cannot be derived within the low energy theory.  Thus, we must
simply assume the net value to be positive if the analysis in this paper is
to be relevant.

There is one exception to this reasoning.  In one circumstance the low-energy
contribution to $\chi$ is divergent, and so dominates the unknown short
distance contribution.  The RPA contribution to $\chi$, which is known as the
Landau diamagnetism, can be put in the form\footnote {This result was brought
to my attention by D. Scalapino. The three dimensional version is given, for
example, in ref.~[25]}
\eq
\chi_{\rm RPA} = \frac{1}{24\pi^2} \oint \frac{dp}{v_F} \Biggl(
\frac{\del^2 {\cal E}}{\del k_x^2} \frac{\del^2 {\cal E}}{\del k_y^2} -
\frac{\del^2 {\cal E}}{\del k_x \del k_y}
\frac{\del^2 {\cal E}}{\del k_x \del k_y}
\Biggr).
\e{landai}
\eqe
The integral runs around the Fermi surface.  A van Hove singularity is a point
where the single particle energy ${\cal E}({\bk})$ has a saddle point, so
${\bv}_F = \nabla {\cal E}$ vanishes linearly.  If the Fermi surface passes
through such a singularity, the integral diverges logarithmically.  Moreover,
the integrand is always negative at a van Hove singularity.  Thus, if the
Fermi surface lies sufficiently close to a van Hove singularity, there is a
large negative contribution to $\chi$ and the uniform phase is unstable.
This is consistent with the mean field studies[17].  These show that at
half-filling with nearest neighbor hopping only (where there is a van Hove
singularity) the uniform phase is not stable, but that introducing doping or
next-nearest neighbor hopping (either of which will move the Fermi level away
from the van Hove singularity) can stabilize the uniform phase. This
discussion is not complete because we have taken the free spinon propagator.
We have not found a form as simple as~\ee{landai} in the full theory.

In conclusion, by the use of the large-$n$ approximation we have been able to
understand some of the physics of the spinon-gauge system.  The main
conclusions are the existence of a nontrivial fixed point and the absence of
weak-coupling antiferromagnetism.  We see no indication of a qualitative
change at small $n$, but cannot exclude it.  For the application to the
high-$T_c$ cuprates, it is off course important to obtain a comparable
understanding of the holons. \\[40pt]

\centerline{\bf Acknowledgements}

I would like to thank B. Blok, G. Minic, M. Natsuume, and D. E. Smith for
helpful conversations, and especially H. Monien and D. Scalapino for their
patient responses to my questions.  This work was supported in part by
National Science Foundation grants PHY89-04035 and PHY90-09850, by the Robert
A. Welch Foundation, and by the Texas Advanced Research Foundation.

\vfill

\pagebreak

\centerline{\bf Figure Captions}

\begin{itemize}

\item[1.] Integral equations for the spinon and
gauge self-energies, assuming Migdal's theorem.  The shading indicates
the full spinon and gauge propagators, while the vertices are uncorrected.
The quartic vertex turns out to be irrelevant.

\item[2.] One-loop vertex correction.

\item[3.] Ladder and crossed-ladder interactions between a spinon (upward
directed arrow) and hole (downward directed arrow), both having momenta near
$\bk_0$.

\item[4.] Renormalization of the local operator ${\cal O} = f^\dagger_\alpha
f_\alpha f^\dagger_\beta f_\beta$.\\
a,b) Exchange in the antiferromagnetic
channel, obtained from fig.~3a by contracting one gauge line.\\
c,d) Exchange
in the BCS channel, obtained from fig.~3b by contracting one gauge line.

\end{itemize}
{\it The figures are not in electronic form, but are quite simple, so I
describe them.  Figure 1 is the obvious one-loop spinon and gauge self-
energies, except that the full propagators appear in the loop.  Figure 2
is again the obvious one-loop vertex correction with full propagators.
Figure 3 shows exchange of two gauge lines between spinon and antispinon;
3a is the ladder and 3b the crossed ladder (full propagators again).
Figure 4 shows the quartic $f f^\dagger f f^\dagger$ vertex, with an
additional gauge line exchanged (a) between initial spinon and antispinon
(b) between final spinon and anti (c) between initial spinon and final anti
(d) between final spinon and initial anti (full propagators).  Note that
exchange between initial and final spinon, or initial and final anti,
is small (nonlogarithmic) as in figure 2.}

\pagebreak

\centerline{\bf References}

\begin{itemize}

\item[1.] R. Shankar, Physica {\bf A177} (1991) 530;
``Renormalization Group Approach to Interacting Fermions,'' Yale preprint
(1992).

\item[2.] J. Polchinski, ``Effective Field Theory and the Fermi Surface,''
ITP preprint NSF-ITP-92-132 (1992), to appear in the Proceedings of the 1992
TASI.

\item[3.] G. Benfatto and G. Gallavoti, J. Stat. Phys. {\bf 59} (1990) 541;
Phys. Rev. {\bf B42} (1990) 9967.

\item[4.] J. Feldman and E. Trubowitz, Helv. Phys. Acta {\bf 63} (1990) 157.

\item[5.] {\it High Temperature Superconductivity, Proceedings of the 1989
Los Alamos Symposium,} ed. K. S. Bedell, et. al., Addison-Wesley, Redwood
City CA, 1990.

\item[6.] P. Anderson and R. Schrieffer, Physics Today (June 1991) 55.

\item[7.] {\it Proceedings of High Temperature Superconductors III,
Kanazawa, Japan, 1991,} in Physica, {\bf C185-C189}.

\item[8.] S. Marten, A. T. Fiory, R. M. Fleming, L. F. Schneemeyer, and J.
V. Waszcak, Phys. Rev. {\bf B41} (1990) 846;\\
P. V. Sastry, J. V. Yakhmi, R. M. Iyer, C. K. Subramanian, and R. Srinivasan,
Physica {\bf C178} (1991) 110.

\item[9.] T. Holstein, R. E. Norton and P. Pincus, Phys. Rev. {\bf B8} (1973)
2649.

\item[10.] E. Fradkin, {\it Field Theories of Condensed Matter Systems,}
Addison-Wesley, Redwood City, 1991.

\item[11.] S. Spielman, J.S. Dodge, L. W. Lombardo, C. B. Eom, M. M. Fejer,
T. H. Geballe, and A. Kapitulnik, Phys. Rev. Lett. {\bf 68} (1992) 3472.

\item[12.] P. A. Lee, Phys. Rev. Lett. {\bf 63} (1989) 680.

\item[13.] L. B. Ioffe and P. B. Wiegmann, Phys. Rev. Lett. {\bf 65} (1990)
653.

\item[14.] L. B. Ioffe and G. Kotliar, Phys. Rev. {\bf B42} (1990) 10348.

\item[15.] N. Nagaosa and P. Lee, Phys. Rev. {\bf B43} (1991) 1233.

\item[16.] G. Baskaran and P. W. Anderson, Phys. Rev. {\bf B37} (1988) 580;\\
P. W. Wiegmann, Phys. Rev. Lett. {\bf 60} (1988) 821.

\item[17.]  L. B. Ioffe and A. I. Larkin, Phys. Rev. {\bf B39} (1989) 8988;\\
J. B. Marston and I. Affleck, Phys. Rev. {\bf B39} (1989) 11538.

\item[18.]  B. Blok and H. Monien, Phys. Rev. {\bf B47} (1993) 3454.

\item[19.] D. E. Smith, in preparation.

\item[20.] L. B. Ioffe and V. Kalmeyer, Phys. Rev. {\bf B44} (1990) 750.

\item[21.] P. W. Anderson, Phys. Rev. Lett. {\bf 65} (1991) 2306; {\bf 66}
(1991) 3226.

\item[22.] M. Yu. Reizer, Phys. Rev. {\bf B39} (1989) 1602;\\
G. Baym, H. Monien, C. J. Pethick, and D. G. Ravenhall, Phys. Rev. Lett.
{\bf 64} (1990) 1867.

\item[23.] B. I. Halperin, P. A. Lee, and N. Read, ``Theory of the
Half-Filled Landau Level,'' preprint (1992).

\item[24.] L. B. Ioffe, S. Kivelson, and A. I. Larkin, Phys. Rev. {\bf B44}
(1991) 12537.

\item[25] R. Peierls, Z. Phys. {\bf 80} (1933) 763.

\end{itemize}

\end{document}